\documentclass[12pt]{amsart}
\usepackage[dvips]{color}
\usepackage{amsmath}
\usepackage{amsxtra}
\usepackage{amscd}
\usepackage{amsthm}
\usepackage{amsfonts}
\usepackage{amssymb}
\usepackage{eucal}
\usepackage{epsfig}
\usepackage{graphics}
\usepackage{epsf,graphics,mathrsfs,yfonts,eufrak,simplewick}
\usepackage{pdfpages}
\usepackage{psfrag}
\usepackage[OT2,T1]{fontenc}

\def\'{\char126}
\def\`{\char127}

\def\({\left(}
\def\){\right)}

\newcommand{\nn}{\nonumber}

\newenvironment{tenumerate}{
  \begin{enumerate}
  
  }{\end{enumerate}}
\newcommand{\bi}{\begin{tenumerate}}
\newcommand{\ei}{\end{tenumerate}}
\newcommand{\isoto}[1][]%
{{\mathop{\buildrel{\sim}\over\longrightarrow}\limits_{#1}}}

\def\[{\left[}
\def\]{\right]}
\newcommand{\la}{\lambda}

\newcommand{\al}{\alpha}

\numberwithin{equation}{section}

\def\half{\textstyle{\frac  1 2}}

\def\bi{\mathbf{i}}

\usepackage[dvipsnames]{xcolor}

\begin{document}
\begin{title}[Suzuki equations and integrals of motion]{
Suzuki equations and integrals of motion for supersymmetric CFT}
\end{title}
\date{\today}
\author{C.~Babenko and  F.~Smirnov}
\address{CB, FS\footnote{Membre du CNRS}: 
 Sorbonne Universite, UPMC Univ Paris 06\\ CNRS, UMR 7589, LPTHE\\F-75005, Paris, France}\email{cbabenko@lpthe.jussieu.fr,smirnov@lpthe.jussieu.fr}

\begin{abstract}

Using equations proposed by J. Suzuki we compute numerically the first three integrals of motion for $N=1$ supersymmetric CFT. 
Our computation  agrees with the results of ODE-CFT correspondence which was explained in a more general context by S. Lukyanov.

\end{abstract}

\maketitle

\section{Introduction}

The present paper contains some preliminary results for a larger project
which consists in computing the one-point functions for the supersymmetric sine-Gordon model (ssG)
generalising the results of \cite{OP,HGSV} obtained for the sine-Gordon case (sG). This problem is interesting because
the integrable description of the space of local operators
for the ssG model should be derived from that of the inhomogeneous 
19-vertex Fateev-Zamolodchikov model while for the sG case it was related to
the inhomogeneous 6-vertex model. 
There is an interesting difference between the two cases: for the 6-vertex case the
local observables are created by two fermions while for the 19-vertex case one has
to introduce additional Kac-Moody current \cite{JMS-FZ}.

The first indispensable step consists in finding the corresponding description in the
conformal case like in the paper \cite{HGSIV}. The generalisation is already not quite
trivial. For example, in the computations of the ground state eigenvalues of the
local integrals of motion the paper   \cite{HGSIV} follows  the procedure
proposed in \cite{BLZII}, namely it uses the Destri-DeVega equations on a half-infinite 
interval. This allows to develop an analytical procedure for the computation of the
eigenvalues in question. Then the procedure is generalised in order
to compute the expectation values on a cylinder of the CFT operators in the fermionic basis. Unfortunately, similar procedure for the super CFT case is
unknown to us, and we are forced to proceed with numerical computations
based on equations which for the 19-vertex model were proposed by J. Suzuki
\cite{suzuki}. It should be said that Suzuki equations have been used already for ssG model and
its conformal limit in \cite{hrs}. 

In the present paper we shall apply the Suzuki equations to the ssG model. 
In the high temperature limit we compute numerically the eigenvalues of
the first three local integrals of motion. We interpolate the results getting
exact general formulae. This way of proceeding may look strange having in mind
that the formulae in question can be alternatively obtained by the ODE-CFT correspondence \cite{DT,BLZ-ODE} following Lukyanov \cite{luk} as will be explained.
However, one should have in mind that we are doing a preliminary work, intending
in future to proceed with similar methods to the one-point functions for which
not much is known. 

The paper is organised as follows. In the first section we give a very brief account
of the ssG model viewed as a perturbed CFT. In the second section we give some
exposition of the Suzuki equations, this is very close to the original work \cite{suzuki}. 
The third section contains numerical results and their interpolation. Finally, in the
last section we explain how the eigenvalues are obtained from ODE-CFT correspondence
following \cite{luk}.

\section{Supersymmetric sine-Gordon model}

We begin with a very brief description of the supersymmetric sine-Gordon (ssG) field theory,
an interested reader can find all necessary details in \cite{BD}.
In the framework of Perturbed CFT (PCFT) the ssG is considered as a perturbation of  of the $c=3/2$ CFT (one boson+one Majorana fermion) by the relevant operator $\Phi=-\mu \bar{\psi}\psi\cos\(\frac{\beta\varphi}{\sqrt{2}}\)$:
\begin{align}
\mathcal{A}=\int\Bigl( \frac 1 {16\pi}\partial_z\varphi\partial_{\bar{z}} \varphi+\frac 1 {2\pi}\(\psi\partial_{\bar z}\psi+\bar{\psi}\partial_z\bar{\psi}\)&-2\mu \bar{\psi}\psi\cos\(\frac{\beta\varphi}{\sqrt{2}}\)\Bigr)d^2z\,.\label{action}%&-\frac{2\pi\mu^2}{\beta^2}\cos\(\sqrt{2}\beta\varphi\)\Bigr)d^2z\,,\nn
\end{align}
%\textcolor{red}{Not sure if matters but Bajnok has different signs for the perturbing terms}
The dimensional coupling constant  $\mu$ is of dimension is $[\mathrm{mass}]^{1-\beta^2}$.
The scaling dimension of this operator $\Delta_\mathrm{pert}=\frac 1 2(1+\beta^2)$ is greater than
$\frac 1 2$, so the UV regularisation is needed. 
The OPE 
\begin{align}\Phi(z,\bar z)\Phi(0)=\frac 1 {(z\bar z)^{1+\beta^2}}+C\cdot \frac 1 {(z\bar z)^{1-\beta^2}}\cos\(\sqrt{2}\beta\varphi\)+\cdots\,,\label{OPE}
\end{align}
shows that the UV regularisation is simple: the first non-trivial contribution comes with integrable singularity. The model is shown to be integrable, actually
this is the simplest example of perturbations of parafermionic models whose integrals of motion are obtained in 
 \cite{fateev}. The factorisable S-matrix is known, it coincides with the S-matrix for the spin-1 integrable magnetic \cite{resh}, in the context of
 relativistic field theory it was discussed in \cite{ahn}. The S-matrix is compatible with the $N=1$ supersymmetry.

The formula for the action  \eqref{action}  may contradict the reader's intuition because the supersymmetric classical action contains the additional
term  $ \Phi_1=-\frac{\pi\mu^2}{\beta^2} \cos\(\sqrt{2}\beta\varphi\)  $ 
%\textcolor{red}{Comparing with Bajnok I would have rather put $\Phi_1=-\frac{\pi\mu^2}{\beta^2} \cos\(\sqrt{2}\beta\varphi\) $ }
which we have seen already in the OPE \eqref{OPE}. In the frame work of 
the PCFT this term, as it is written, cannot be added to the action for dimensional reasons, at least it needs a new dimensional coupling constant. In the
classical limit $\beta\to 0$ the situation becomes more complicated. That is why, when proceeding in the opposite direction, i.e. quantising the classical
model
by more traditional methods of QFT, one should indeed begin with the supersymmetric action which includes $\Phi_1$ and take care of preserving the supersymmetry.
This was done in \cite{lm}, the result is exactly as expected from our dimensional considerations: the dimensional coupling constants
for the two terms of the interaction are renormalised differently, the term with  $\Phi_1$ containing vanishing power of the cutoff. 

Like in the sine-Gordon case it is often convenient to rewrite the action as
\begin{align}
\mathcal{A}=
\int\Bigl[\Bigl( \frac 1 {16\pi}\partial_z\varphi\partial_{\bar{z}} \varphi+\frac 1 {2\pi}\(\psi\partial_{\bar z}\psi+\bar{\psi}\partial_z\bar{\psi}\)&-\mu \bar{\psi}\psi e^{-i\frac{\beta}{\sqrt{2}}\varphi}\Bigr)-\mu \bar{\psi}\psi e^{i\frac{\beta}{\sqrt{2}}\varphi}\Bigr]d^2z\,,\label{action1}
\end{align}
considering the model as perturbation of a supersymmetric CFT with the Virasoro central charge
equal to $c=\frac 3 2 \(1-2(\beta-\beta^{-1})^2\)$,  by the relevant operator (the last term) with scaling dimension $\Delta=\beta^2$.

The mass of the fundamental particles is exactly related to the dimensional coupling constant by
a formula of Al. Zamolodchikov's type 
\begin{align}
M=\frac{4(1-\beta^2)}{\pi \beta^2}\(\frac{\pi} 2\mu \gamma\Bigl(\frac {1-\beta^2} 2\Bigr)\)^{\frac 1 {1-\beta^2}}\,,\label{mass}
\end{align}
where $\gamma(x)=\Gamma(x)/\Gamma(1-x)$.

\section{Suzuki equations}

In this section we shall use more appropriate parameters for the lattice case :
$$\nu =\half(1-\beta^2)\,\qquad q=e^{\pi i \nu}\,.$$

Consider an inhomogeneous XXZ chain of spin $1$ of even length $L$ with  twist $q^{\kappa}$. In order to avoid multiple change of variables we shall work from the very beginning  with the rapidity-like ones. The relation to usual
multiplicative variables $\la$ \cite{JMS-FZ} is $\la =e^{\pi i \nu\theta}$. 
We shall consider two transfer-matrices  corresponding  to auxiliary spaces of spins $1/2$ and $1$. Corresponding ground state eigenvalues will be
denoted respectively by $T_1(\theta)$, $T_2(\theta)$.

The Baxter equations take the form
\begin{align}
T_1(\theta)Q(\theta)=a(\theta)Q(\theta+\pi i)+d(\theta)Q(\theta-\pi i)\,,
\end{align}
where $a(\theta)$ and $d(\theta)$ are trigonometric polynomials:
$$a(\theta)=\prod_{j=1}^L\sinh\nu(\theta-\tau_j-\pi  i),\quad d(\theta)=\prod_{j=1}^L\sinh\nu(\theta-\tau_j+\pi  i)\,,$$
%\textcolor{red}{maybe change here $L$ to another letter since $L$ will be one of the kernels}

$T_1(\theta)$ is a trigonometric polynomial of the same form and the same degree, finally
\begin{align}Q(\theta)=e^{\nu\kappa\theta}\prod_{j=1}^m\sinh\nu(\theta-\sigma_j)\,,\label{Q}\end{align}
$\sigma_j$ being the Bethe roots. We shall be interested in the case of real $\tau_j$ and $\kappa$ which implies
$$\overline{a(\theta)}=d(\bar{\theta})$$

We are interested in the ground state for which $m=L$. For  large $L$ and sufficiently small $\kappa$ the Bethe roots are close
to the two-strings: $\sigma_{2j-1}\simeq \eta_j-\pi i/2, \sigma_{2j}\simeq \eta_j+\pi i/2$ for certain real $\eta_j$. 

The transfer-matrix $T_2(\theta)$ is obtained by the fusion relation:
\begin{align}T_2(\theta)=T_1(\theta-\pi i /2)T_1(\theta+\pi i /2)-f(\theta)\,,\label{fusion}\end{align}
here and later
$$f(\theta)=a(\theta-\pi i/2)d(\theta+\pi i/2)\,.$$
According to the investigation done by Suzuki \cite{suzuki} the zeros of $T_1(\theta)$ lie approximately on the lines $\mathrm{Im}(\theta)=\pm 3\pi i /2$, and
zeros of $T_2(\theta)$ lie approximately on the lines $\mathrm{Im}(\theta)=\pm \pi i $, $\mathrm{Im}(\theta)=\pm 2\pi i $.

Let us introduce the auxiliary functions 
\begin{align}y(\theta)=\frac{T_2(\theta)}{f(\theta)},\quad Y(\theta)=1+y(\theta)\,.\label{defy}\end{align}
The function $\log(T_2(\theta))$  grows for $\mathrm{Re}(\theta)\to\pm\infty$  slowly (as $\pm 2L\theta$). This allows to derive from
\eqref{fusion} the first important relation:
\begin{align}
\log T_1(\theta)=(L*\log(f Y))(\theta)
%\int\limits_{-\infty}^{\infty}L(\theta-\theta')\log\(f(\theta')Y(\theta')\)d\theta'
\,,\label{1eq}
\end{align}
where we introduced the kernel which will be often used:
$$L(\theta)=\frac {1}{2\pi \cosh\theta}\,.$$
and $\ast$ means the usual convolution product.

We have
$$T_2(\theta)=\lambda_1(\theta)+\lambda_2(\theta)+\lambda_3(\theta)\,,$$
where
\begin{align}
&\lambda_1(\theta)=a(\theta+\pi i /2)a(\theta-\pi i /2)\frac{Q(\theta+3\pi i /2)}{Q(\theta-\pi i /2)}\,\nn\\
&\lambda_2(\theta)=a(\theta+\pi i /2)d(\theta-\pi i /2)\frac{Q(\theta-3\pi i /2)Q(\theta+3\pi i /2)}{Q(\theta-\pi i /2)Q(\theta+\pi i /2)}\,\nn\\
&\lambda_3(\theta)=d(\theta+\pi i /2)d(\theta-\pi i /2)\frac{Q(\theta-3\pi i /2)}{Q(\theta+\pi i /2)}\,.\nn
\end{align}
The second  auxiliary function is defined by
\begin{align}
b(\theta)=\frac{\la_1(\theta+\pi i /2)+\la_2(\theta+\pi i /2)}{\la_3(\theta+\pi i /2)}\,,\quad B(\theta)=1+b(\theta)\,.
\end{align}
Using the Baxter equation we derive
\begin{align}
b(\theta)=T_1(\theta)\frac{Q(\theta+2\pi i)}{Q(\theta-\pi i)}\frac{a(\theta+\pi i)}{d(\theta)d(\theta+\pi i)}\,.\label{bQ}
\end{align}
On the other hand it is obvious from the definition that
\begin{align}
T_2(\theta+\pi i /2)=B(\theta)d(\theta+\pi i)d(\theta)\frac{Q(\theta-\pi i )}{Q(\theta+\pi i )}\,.\label{aux1}
\end{align}
Multiplying the latter equation by the conjugated one for real $\theta$ one easily derives
the second important equation
\begin{align}
\log y(\theta)=(L * \log(B \overline{B}))(\theta)
%\int\limits _{-\infty}^{\infty}L(\theta-\theta')\log(B(\theta')\overline{B}(\theta'))d\theta'
\,.\label{2eq}
\end{align}

Now comes the main of Suzuki's tricks. Consider a function $G(\theta)$ which is regular in the strip $0<\mathrm{Im}(\theta)<\pi$, and which decrease sufficiently
fast at $\pm\infty$. Then having in mind the structure of zeros of $T_2(\theta)$ described above we have
\begin{align}
\int\limits _{-\infty}^{\infty}\(G(\theta-\theta')
%\frac{d}{d\theta'}
\log T_2(\theta'+\pi i/2)-G(\theta-\theta'+\pi i )
%\frac{d}{d\theta'}
\log T_2(\theta'-\pi i/2)
\)d\theta'=0\,.\label{aux}
\end{align}
Using \eqref{aux1} we rewrite this as follows
\begin{align}
&\int\limits_{-\infty}^{\infty}\(G(\theta-\theta')+G(\theta-\theta'+\pi i )\)\log\frac{Q(\theta'+\pi i )}{Q(\theta'-\pi i )}d\theta'\nn\\&=
\int\limits_{-\infty}^{\infty}\(G(\theta-\theta')\log(d(\theta')d(\theta'+\pi i ))-G(\theta-\theta'+\pi i )\log(a(\theta')a(\theta'-\pi i ))\)d\theta'\nn\\&+
\int\limits_{-\infty}^{\infty}\(G(\theta-\theta')\log(B(\theta'))-G(\theta-\theta'+\pi i )\log(\overline{B}(\theta'))\)d\theta'\,.\nn
\end{align}
The goal now is to rewrite the left hand side in terms of the auxiliary function $y(\theta),b(\theta)$. 
From \eqref{bQ} and \eqref{1eq} one derives
\begin{align}
\log b(\theta)=\log\(\frac{Q(\theta+2\pi i)}{Q(\theta-\pi i)}\)+\log\(\frac{a(\theta+\pi i-i0)}{d(\theta)d(\theta+\pi i)}\)
+
\int\limits_{-\infty}^{\infty}L(\theta-\theta')\log\(f(\theta')Y(\theta')\)d\theta'\,.\nn
\end{align}
So, our goal will be achieved if we find such $G(\theta)$ that
\begin{align}\int\limits_{-\infty}^{\infty}\(G(\theta-\theta')+G(\theta-\theta'+\pi i )\)\log\frac{Q(\theta'+\pi i )}{Q(\theta'-\pi i )}&=
\log\(\frac{Q(\theta+2\pi i)}{Q(\theta-\pi i)}\)\nn\\&+\pi i \nu\kappa(4 {G}_0-3)\,,
 \end{align}
where the last term takes account of the the multiplier $e^{\nu\kappa \theta}$ in $Q(\theta)$, ${G}_0$ being the average of $G$ over the real line.
Recalling that in the formula for $Q(\theta)$ \eqref{Q} the Bethe roots are approximately two-string one easily finds
$G(\theta)$ by Fourier transform:
\begin{align}
G(\theta)=\frac 1 {4\pi}\int\limits _{-\infty}^{\infty}\frac{\sinh\(\frac{\pi k}{2\nu}(1-3\nu)\)}{\sinh\(\frac{\pi k}{2\nu}(1-2\nu)\)\cosh\(\frac{\pi k}{2}\)}e^{-ik \theta}dk\,.\label{G}
\end{align}
Notice that ${G}_0=\frac{1-3\nu}{2(1-2\nu)}$.

Finally, after some computation we arrive at
\begin{align}
\log b(\theta)&=2\sum_j\log\Bigl(\tanh\frac 1 2(\theta-\tau_j-i0)\Bigr)%-2i\sum_j\arctan\Bigl(\tanh\frac 1 2(\theta-\tau_j)\Bigr)
-\frac{\pi i \nu\kappa}{1-2\nu}\label{eqb}\\&+(L*\log Y)(\theta)+(G*\log B)(\theta)-(G*\log\overline{B})(\theta+\pi i)\,.\nn
\end{align}

We obtain the massive relativistic model from the inhomogeneous lattice one by the usual prescription: set $\tau_j=(-1)^j\tau$ and consider the limit
$$\tau\to \infty\,,\quad L\to \infty,\quad 2L  e^{-\tau}\to 2\pi MR\,\,\,\, \text{finite}\,.$$
In this limit
$$2\sum_j\log\Bigl(\tanh\frac 1 2(\theta-\tau_j)\Bigr)\to-2\pi MR\cosh(\theta)\,.$$

The idea is that in this limit we should obtain the eigenvalue of the %\textcolor{red}{field theoretic}
 transfer-matrix corresponding to the NS ground state
with the twist defined by 
\begin{align}
 \sqrt{2} \beta P=\nu\kappa\,.\label{defP}
\end{align}
Here $\sqrt{2}$ comes from the normalisation of the topological charge consistent with \eqref{action}.
The normalisation of this twist is explained by the requirement that in the high temperature limit $R\to 0$ the eigenvalue of the
first integral of motion, $I_1$, which is nothing but $L_0-c/24$ is given by
$${i}_1=P^2  -\frac 1 {16}\,.$$

\section{Numerical work}

The function $b(\theta)$ rapidly decreases when $\mathrm{Re}(\theta)\to\pm\infty$, $0>\mathrm{Im}\theta>-\pi/2$. Introducing the shift $0<\pi\gamma<\pi/2$ and
moving the contours of integration we arrive at the system which allows a numerical investigation:
\begin{align}
&\log b(\theta-\pi i\gamma)=-2\pi MR\cosh(\theta-\pi i \gamma)
-\frac{\pi i \sqrt{2}}{\beta}P+\half\log 2\label{eqbfinal}\\&+\int\limits_{-\infty}^{\infty}L(\theta-\theta'+\pi i \gamma)
\log\(\half Y(\theta')\)d\theta' %\(\frac{\pi} 2\mu \gamma\Bigl(\frac {1-\beta^2} 2\Bigr)\)^{\frac 1 {1-\beta^2}}
\nn\\&+\int\limits_{-\infty}^{\infty}\Bigl[G(\theta-\theta')\log B(\theta'-\pi i \gamma)-G(\theta-\theta'+\pi i (1-2\gamma))\log\overline{B(\theta'-\pi i \gamma)}\Bigr]d\theta'\,.\nn\\
&\log y(\theta)=\int\limits_{-\infty}^{\infty}2\mathrm{Re}\Bigl[L(\theta-\theta'+\pi i \gamma)\log B(\theta'-\pi i \gamma)]
%+L(\theta-\theta'-\pi i \gamma)\log\overline{B(\theta'-\pi i \gamma)}\Bigr]
d\theta'\,.\label{eqyfinal}
\end{align}

%\textcolor{red}{I suppressed extra factor $(\frac{\pi} 2\mu \gamma\Bigl(\frac {1-\beta^2} 2\Bigr) )^{\frac 1 {1-\beta^2}} $ appearing in the equation below}

The integrals containing $\log B$ converge at infinities very fast because the absolute value of the integrand
is estimated as $\exp (-Const \cdot e^{|\theta|})$ with positive $Const$. The integral with $\log (\half Y)$ converges much more slower
because $y(\theta)$ behaves as $1+O(e^{-|\theta|})$. In the numerical computations we replace integrals by finite sums, and the above
estimates mean that the number of points needed for the approximation of the integral containing $\log (\half Y)$ should be bigger than that for the integrals
containing $\log B$.

Our goal is to consider the high temperature limit $R\to0$. The previous formulae are simplified if we
use the parametrisation:
\begin{equation}
R=\frac {\beta}{\sqrt{2}}\(\frac{\pi} 2\mu \gamma\Bigl(\frac {1-\beta^2} 2\Bigr)\)^{-\frac 1 {1-\beta^2}}e^{-\theta_0}\,,
\label{defR}
\end{equation}
with $\theta_0$ being a dimensionless parameter. Now the driving term in the equation \eqref{eqbfinal}becomes
$$-4\sqrt{2}\frac {1-\beta^2}\beta e^{-\theta_0}\cosh(\theta-i\gamma)\,.$$

The  local integrals of motion are extracted form $y(\theta)$ which is the normalised transfer-matrix of auxiliary spin 1 \eqref{defy}.
Namely, for $\theta\to\infty$ the asymptotical formula holds:
\begin{align}
\log y(\theta)\simeq \sum\limits_{k=1}^{\infty}C_{2k-1}i_{2k-1}(\theta_0)e^{-(2k-1)\theta}\,,\label{asy}
\end{align}
similarly the asymptotics for  $\theta \to-\infty$ is related to $\bar i_{2k-1}(x)$.
The constants $C_m$ are given by
\begin{align}
C_m=-\frac{\beta}{\sqrt{2}(1-\beta^2)}\frac{\sqrt{\pi}\ \Gamma\(\frac m 2 \)\Gamma\(\frac 1 {1-\beta^2}m\)}{(m-1)!\(\frac{m+1}2\)!\Gamma\(1+\frac {\beta^2}{1-\beta^{2}}m\)}\,.
\label{defC}
\end{align}

%\textcolor{red}{I changed $(m-1)!$ to $m!$ in $C_m$  }

%\textcolor{blue}{Strange}

This  normalisation is chosen for the sake of the conformal limit, the appearance of this kind of coefficients is not surprising for a reader
familiar with \cite{BLZII}, we shall give more explanation in the next section.

The main advantage of the above normalisation is that in the high temperature limit we have
$$e^{-(2k-1)\theta_0}i_{2k-1}(\theta_0)\ \ \to\hskip -.7cm\raisebox{-.2cm}{${}_{\theta_0\to\infty}$}\ \ \  i_{2k-1}\,,$$
with $ i_{2k-1}$ being the local integrals of motion for the CFT case normalised as follows:
$$i_{2k-1}=P^{2k}+\cdots\,.$$

Now we start the numerical work. Our goal is to obtain the formulae for $i_1,i_3,i_5$ by interpolation in $P$ and $\nu$. 
This may sound as a purely academic exercise having in mind that these formulae can be obtained analytically
as explained in the next section. However, in our further study we shall need to guess the formulae for the
one-point functions in the integrable basis of supersymmetric CFT, which are unknown. That is why we want to be sure that
our numerical methods are sufficiently precise. 

The twist $P$ cannot be too large, we restrict ourselves to $P\le 0.2$, 
%\textcolor{red}{I am not sure  if it matters but it is the old $p$ range, for $P$ the should a priori be modified depending on the value of $\nu$ }
 we take $\beta$ sufficiently close to $1$.
For given $\beta$ we
interpolate in $P$ from the solutions to \eqref{eqbfinal}, \eqref{eqyfinal} for $\theta_0=18$. 
Integrals are replace by sums with  step $0.1$, the shift is $\gamma=0.1$, the limits in the integrals containing $\log B(\theta-\pi i \gamma)$
are $[-24, 24]$, the limits of the integral containing $\log (Y(\theta)/2)$ are $[-72, 72]$.

We  normalise by the leading coefficient which is later compared with $C_{2k-1}$. 
Doing that
for a sufficient  number of different $\beta$'s and assuming that due to the general structure of CFT the local integrals must be polynomials in 
$$Q^2=-\frac{(1-\beta^2)^2}{\beta^2}\,,$$ 
we were able to
interpolate further:
\begin{align}
&i_1=P^2-\frac 1 {16}\,\label{egint}\\
&i_3=P^4-\frac 5 {16}P^2+\frac 1 {512}(9+2Q^2)\,,\nn\\
&i_5=P^6-\frac{35}{48}P^4+\frac{537 + 46 Q^2}{3072}P^2-\frac{475 +190 Q^2 + 24 Q^4}{49152}\,.\nn
\end{align}
\vskip.3cm

%\textcolor{red}{I changed the values of the coefficents in $i_{5} $ (new values checked both with semiclassics and numerics)}

We shall not go into the details of the interpolation restricting ourselves  to two examples
in which we compare the results of the numerical computations using the equations \eqref{eqbfinal}, \eqref{eqyfinal} with the
analytical formulae
 \eqref{defC}, \eqref{egint}.

It is more direct to compare computational results with
$$j_{m}=C_mi_m\,.$$

Here are the results for $\beta^2=\frac 1 2$:

%\textcolor{red}{I replaced the old $p$ notation in the tables to $P$ }

\vskip .3cm

\scalebox{.9}{\begin{tabular}{|r|r|r|r|r|r|r|}
  \hline
  $P$ & $j_1$ comp. & $j_3$ comp. & $j_5$ comp. & $j_1$ analyt.& $j_3$ analyt.& $j_5$ analyt.\\
  \hline
0.02& 0.195092899& -0.121737971& 0.385270717&0.195092904& -0.121737972& 0.385270720
\\
\hline
0.04& 0.191322988& -0.118811577& 0.375422434& 0.191322993& -0.118811578& 0.375422438\\
\hline
0.06& 0.185039803& -0.113984520& 0.359237764& 0.185039807& -0.113984521& 0.359237767
\\
\hline
0.08& 0.176243343& -0.107332198& 0.337056416& 0.176243348& -0.107332199& 0.337056419
\\
\hline
0.1& 0.164933610& -0.0989601675& 0.309346006& 0.164933614& -0.0989601686& 0.309346008
\\
\hline
0.12& 0.151110603& -0.0890041464& 0.276694070& 0.151110607& -0.0890041473& 0.276694072
\\
\hline
0.14& 0.134774321& -0.0776300103& 0.239797812& 0.134774325& -0.0776300111& 0.239797814
\\
\hline
0.16& 0.115924766& -0.0650337947& 0.199451558& 0.115924769& -0.0650337954& 0.199451559
\\
\hline
0.18& 0.0945619364& -0.0514416943& 0.156531934& 0.0945619389& -0.0514416947& 0.156531935
\\
\hline
0.2& 0.0706858328& -0.0371100629& 0.111980775& 0.0706858347& -0.0371100632& 0.111980775
\\
\hline

\end{tabular}}
\newpage

Here are the results for $\beta^2=\frac 3 5$:

\vskip .3cm

\scalebox{.9}{\begin{tabular}{|r|r|r|r|r|r|r|}
  \hline
  $P$ & $j_1$ comp. & $j_3$ comp. & $j_5$ comp. & $j_1$ analyt.& $j_3$ analyt.& $j_5$ analyt.\\
  \hline
0.02& 0.267141860& -0.315491660& 1.87869822& 
  0.267141961& -0.315491728& 1.87869854\\ \hline0.04& 
  0.261979700& -0.308328920& 1.83430033& 0.261979797& -0.308328984& 
  1.83430063\\ \hline0.06& 0.253376100& -0.296514050& 1.76131551& 
  0.253376191& -0.296514109& 1.76131579\\ \hline0.08& 
  0.241331061& -0.280231598& 1.66124365& 0.241331143& -0.280231651& 
  1.66124390\\ \hline0.1& 0.225844581& -0.259739931& 1.53614935& 
  0.225844653& -0.259739975& 1.53614955\\ \hline0.12& 
  0.206916661& -0.235371233& 1.38862669& 0.206916720& -0.235371267& 
  1.38862685\\ \hline0.14& 0.184547302& -0.207531507& 1.22175395& 
  0.184547345& -0.207531531& 1.22175405\\ \hline0.16& 
  0.158736503& -0.176700578& 1.03903821& 0.158736527& -0.176700590& 
  1.03903826\\ \hline0.18& 0.129484265& -0.143432085& 0.844349942& 
  0.129484268& -0.143432087& 0.844349948\\ \hline0.2& 
  0.0967905868& -0.108353490& 0.641847507& 0.0967905654& -0.108353481&
   0.641847468
   \\
   \hline

\end{tabular}}

\vskip .3cm

%\textcolor{red}{I recover your results for both tables but I use the  coefficients $C_{m}$ and $i_{5}$ that I have put above in comments}

It is clear from these tables that the agreement is quite good. It can be made better by 
choosing bigger $\theta_{0}$, using finer discretisation {\it etc}. But this is not needed for our goals since our precision was sufficient for 
a successful interpolation.

\section{Eigenvalues of integrals from ODE- CFT correspondence}

%\textcolor{red}{There are some signs discrepancies in this section. I wrote corrections considering \eqref{luk} correct.
%However when  I derived \eqref{luk} I got different signs}

The ODE- CFT correspondence is the statement that in the conformal case
the vacuum eigenvalues of the operator $Q(\theta )$ coincide with determinants of certain
ordinary differential equations. 
The eigenvalues of the transfer-matrices $T_j(\theta)$ coincide with certain Stokes multipliers for the corresponding equation.
In the case of $c<1$ CFT this statement goes back to a remarkable observation due to Dorey ans Tateo \cite{DT},
which was later essentially clarified and generalised by Bazhanov, Lukyanov, Zamolodchikov \cite{BLZ-ODE}. We shall not go into details
of further generalisation of the ODE- CFT correspondence and its generalisation to the massive case, restricting ourselves to the case of 
supersymmetric CFT which is considered in the present paper. 
It is useful to consider more general situation of a parafermion $\Psi_k$ interacting with  a free boson because there is
certain difference between $k$ even or odd. The $c=1$ CFT corresponds to $k=1$, and the $c=3/2$ case, considered in
this paper, corresponds to $k=2$.
In general case Lukyanov \cite{luk} proved that
the operator $Q(\theta )$ is related to the following ODE:
\begin{align}
\psi''(z)-\Bigl((z^{2\al}-E)^k+\frac{l(l+1)}{z^2}
\Bigr)\psi(z)=0\,,\label{luk}
\end{align}
%\textcolor{red}{I got different signs : $ \psi''(x)-a^{-2}\Bigl((x^{2\al}+1)^k+\frac{l(l+1)}{z^2}
%\Bigr)\psi(z)=0 $, as written here the sign does not match with sign of the term $x^{-2}S $ in Riccati equation }
the relation of $E,\al,l$ to parameters  $\theta,\beta^2,k,P$ is as follows
\begin{align}
\al=\frac{1-\beta^2}{k\beta^2}\,,\ \ E=\frac \beta{\sqrt{k}}e^{\frac{1-\beta^2}k(\theta-\theta_0)}\,,\ \ l=\frac{\sqrt{k}}\beta P-\frac1  2\,.\label{idn}
\end{align}
and $\theta_{0}$ is defined by a fromula analogous to \eqref{defR}.
The parameter $\al$ is positive, so, we are dealing with a self-adjoint operator on the positive half-line.
Then $Q(E)$ is just its determinant (here and later we allow ourselves to use both $Q(\theta)$ and $Q(E)$
having in mind the identification \eqref{idn}).

The eigenvalues 
$Q(E)$ and   $T_j(E)$ are entire functions of $E$.
We are interested  in their large $E$ asymptotics. It
is known that for $\log Q(E)$ and for  $\log T_j(E)$ with $j$ up to $k-1$ the asymptotics go in
two kinds of exponents: $E^{-\frac {2j-1} { 2k(1-\beta^2)}}$ and $E^{\frac{j}{k\beta^2}}$, ($j\ge1$), the coefficients
being proportional to the eigenvalues of local and non-local integrals of motion. The latter are of no interest for us, 
that is why we shall deal directly with $\log T_k(E)$ which possesses an exceptional property of
containing in its asymptotics $E^{-\frac {2j-1} { 2k(1-\beta^2)}}$ only. In order to explain that we have to
consider \eqref{luk} as an equation of a complex variable.

Let $z=|z|e^{i\varphi}$. Since the parameter $\al$ is generally irrational we are dealing with an infinite covering of the plane:
$-\infty < \varphi<  \infty$. 
%\textcolor{red}{I changed the determination according to further contour integration} 

The main property allowing to investigate the determinant and the Stokes multipliers is that
for any solution $\psi(z,E)$ the function
$$(\Omega\psi)(z,E)=q^{1/2}\psi( pz,q^2E)\,; \quad p=e^{\pi i \beta^2}\,,\ \ q=e^{\pi i \frac {1-\beta^2} k}\,,$$
is also a solution.

Consider the solution $\chi(z,E)$ characterised by the following asymptotics for  real $z\to+\infty$:
$$\chi(z,E)\simeq x^{-\frac{\al k}2}\exp\Bigl(-\frac{x^{\al k+1}}{\al k+1}\Bigr)\,.$$
Following the \cite{BLZ-ODE,luk} and using the fusion relations it is not hard to derive for any $j$ the relation
between the three solutions:
\begin{align}
(\Omega^{j+1}\chi)(z,E)=-T_{j-1}(Eq^{j+1})\chi(z,E)+T_j(E q^{j})(\Omega\chi)(z,E)\,.\nn
\end{align}
The asymptotic behaviour at $E\to \infty$ is investigated by WKB method, where the important role is played by the
the function $\sqrt{(x^\al-E)^k+\frac{l(l+1)}{x^2}}$.% \textcolor{red}{ $\sqrt{(x^{2\al}-E)^k-\frac{l(l+1)}{x^2}}$}.

 One rescales $x$ for  large $E$ so that 
the term $\frac{l(l+1)}{x^2}$ is small.  It is clear that exactly for $j=k$ the function $T_k(E q^{k})$ can be
considered as the Stokes multiplier 
between growing solutions $(\Omega\chi)(z,E)$ and $(\Omega^{k+1}\chi)(z,E)$
for two neighbouring sectors which are semi-classically separated by the cut of the square root. 
%Indeed,  the argument of $p^{2k\al}=e^{2\pi i (1-\beta^2)}$ is less that $2\pi$.
This implies a simple formula for the asymptotics of $\log T_k(E q^{k})$ 
given below.

Let us change variables rewriting \eqref{luk} as
\begin{align}
a^2\psi''(x)-\Bigl((x^{2\al}-1)^k+a^2\frac{l(l+1)}{x^2}
\Bigr)\psi(x)=0\,, \label{luk2}
\end{align}
%\textcolor{red}{$a^2\psi''(x)-\Bigl((x^{2\al}-1)^k-a^2\frac{l(l+1)}{x^2}
%\Bigr)\psi(x)=0$ since the factor $\frac{1}{4x^{2}}$ does not appear }
where $a^2=E^{-\frac k {(1-\beta^2)}}$. %\textcolor{red}{ $a^2=E^{-\frac{k} {(1-\beta^2)}}  $ } 

We prefer to write the WKB formulae in a somewhat XIX century way in order
to avoid some total derivatives. Namely, we present the solution  to \eqref{luk2} in the form
$$\psi(x,x_0)=S(x,a)^{\frac 1 2}\exp\Bigl(\frac 1 a \int\limits_{x_0}^{x}\frac{dy}{S(y,a)}
\Bigr)\,,$$
where $S(x,a)$ satisfies the Riccati equation (we omit arguments)
$$\frac 4 {a^2}\(1-FS^2\)-S'{}^2+2S''S+x^{-2}S^2=0\,,$$ 
%\textcolor{red}{with \eqref{luk2} it is $ \frac{4}{a^2}\(1-FS^2 \)-S'{}^2+2S''S-x^{-2}S^2=0 $ }
with 
$$F(x,a,b)=(x^{2\al}-1)^k+\frac{b^2}{x^2}\,,$$
where we introduce $b=a(l+1/2)$, in spite of the fact that $b\ll 1$ it is convenient to develop into series in this parameter only at the
final stage. 
The ansatz for $\psi$ is different from usual quantum mechanical formulae, and it allows to avoid appearance of redundant total derivatives. 
%The asymptotics of $Q(a,l)$
% \textcolor{red}{$D(a,l)$ not defined, maybe  $Q$ or $ T_{k} $ instead }
%for $a\to\infty$ is obtained as properly regularised $\psi(\infty,0)$. 
Using Riccati  
equation we find for $S(x,a)$ the power
series
\begin{align}S(x,a,b)=\sum_{k=0}^\infty a^{2k}S_k(x,b)\,.\label{seriesS}\end{align}
In particular,
$$
\frac 1 {S_0(x,b)}=\sqrt{F(x,a,b)}\,.
$$

According to our reasoning concerning the Stokes multiplier, we have for the asymptotics
\begin{align}\log T_k(E q^{k})\simeq \frac 1 a \int_C\frac{dy}{S(y,a)}\,,\label{logT}\end{align}
where the contour $C$ goes from $\infty\cdot e^{+i0}$ to  $\infty\cdot e^{-i0}$ around the cut of $\sqrt{F(x,a,b)}$.
Let us consider the contribution from $S_0(x,b)$. Recalling that $b\ll 1$ we develop
\begin{align}
\frac 1 {S_0(x,b)}=\sum\limits_{p=0}^\infty \binom{1/2}{p}(x^{2\al}-1)^{\frac {k(1-2p)} 2}b^{2p}x^{-2p}\,.\nn
\end{align}

%\textcolor{red}{with the new version of $F$ one should add an extra $(-1)^p$ in the sum}

Now the difference between $k$ odd or even becomes clear. %For
%odd $k$ we have
We have to evaluate the integral
$$\int\limits_C(y^{2\al}-1)^{\frac {k(1-2p)} 2}y^{-2p}dy\,.
%=\frac 1 \al B\Bigr(1-\frac{2p-1} 2 k\ ,\ \bigl( \al^{-1}+k\bigr)\frac{2p-1} 2\Bigr)\,,
$$
%while for even $k$ the cut disappears and the integral s computed by residue
%$$\int\limits_C(y^{2\al}-1)^{\frac {k(1-2p)} 2}y^{-2p}dy=\frac{\pi i }{\al}\binom{( \al^{-1}+k)\frac{2p-1} 2-1}{\frac{2p-1} 2 k-1}\,.$$
By the change of variables $w=y^{2\al}$ this integral reduces 
for odd $k$  to a beta-function  and for even $k$ to a binomial coefficient.
In spite of this computational difference the final result does not depend on the parity of $k$, after some simplification we get
$$\int\limits_C(y^{2\al}-1)^{\frac {k(1-2p)} 2}y^{-2p}dy
=\frac{\pi i k \beta^2}{1-\beta^2}e^{-\frac {\pi i} 2k  (2p-1)}\frac
{\Gamma\(\frac{k(2p-1)}{2(1-\beta^2)}\)
}{
\Gamma\(1+\frac{k\beta^2(2p-1)}{2(1-\beta^2)}\)\Gamma\(\frac{ k (2p-1)} 2 \)
}
\,.$$
Plugging this into \eqref{logT} we find the constants $C_m$.
Higher corrections in $a^2$ following from \eqref{seriesS} are considered similarly. For $k=2$ one finds  exactly the expressions 
\eqref{egint}. 

%%%%%%%%%%%%%%%%%%%%%%%%%%%%%%%%%%%%%%%%%%%%%%%%%%%%%%%%%%%%%%%%%%%%%%%%%%%%%%%%%%%%%%%%%%%%%%%%%%%%%%%%%%%%%%%%%%%%%%%%%%%%%%%%%%%%%%%%%%%%%%%%%%%%%%%%%%%%%%%%%%%%%%%%%%%%%%%%%%%%%%%%%%%%%%%%%%%%%%%%%%%%%%%%%%%%%%%%%%%%%%%%%%%%%%%%

\end{document}